\documentclass[12pt]{article}
\usepackage{epsfig}

\newcommand{\mrm}[1]{\mathrm{#1}}

\newcommand{\alphaem}{\alpha_{\mrm{em}}}

\renewcommand{\c}{\mrm{c}}
\renewcommand{\d}{\mrm{d}}
\newcommand{\e}{\mrm{e}}

\newcommand{\g}{\mrm{g}}

\newcommand{\p}{\mrm{p}}
\newcommand{\q}{\mrm{q}}

\renewcommand{\u}{\mrm{u}}

\newcommand{\cbar}{\overline{\mrm{c}}}
\newcommand{\dbar}{\overline{\mrm{d}}}

\newcommand{\qbar}{\overline{\mrm{q}}}

\newcommand{\ubar}{\overline{\mrm{u}}}

\newenvironment{Itemize}{\begin{list}{$\bullet$}%
{\setlength{\topsep}{0.2mm}\setlength{\partopsep}{0.2mm}%
\setlength{\itemsep}{0.2mm}\setlength{\parsep}{0.2mm}}}%
{\end{list}}
\newcounter{enumct}

\newlength{\captivewidth}
\newlength{\abstwidth}
\def\ap#1#2#3   {{\em Ann. Phys. (NY)} {\bf#1} (#2) #3.}
\def\apj#1#2#3  {{\em Astrophys. J.} {\bf#1} (#2) #3.}
\def\apjl#1#2#3 {{\em Astrophys. J. Lett.} {\bf#1} (#2) #3.}
\def\app#1#2#3  {{\em Acta. Phys. Pol.} {\bf#1} (#2) #3.}
\def\ar#1#2#3   {{\em Ann. Rev. Nucl. Part. Sci.} {\bf#1} (#2) #3.}
\def\cpc#1#2#3  {{\em Computer Phys. Comm.} {\bf#1} (#2) #3.}
\def\err#1#2#3  {{\it Erratum} {\bf#1} (#2) #3.}
\def\ib#1#2#3   {{\it ibid.} {\bf#1} (#2) #3.}
\def\jmp#1#2#3  {{\em J. Math. Phys.} {\bf#1} (#2) #3.}
\def\ijmp#1#2#3 {{\em Int. J. Mod. Phys.} {\bf#1} (#2) #3.}
\def\jetp#1#2#3 {{\em JETP Lett.} {\bf#1} (#2) #3.}
\def\jpg#1#2#3  {{\em J. Phys. G.} {\bf#1} (#2) #3.}
\def\mpl#1#2#3  {{\em Mod. Phys. Lett.} {\bf#1} (#2) #3.}
\def\nat#1#2#3  {{\em Nature (London)} {\bf#1} (#2) #3.}
\def\nc#1#2#3   {{\em Nuovo Cim.} {\bf#1} (#2) #3.}
\def\nim#1#2#3  {{\em Nucl. Instr. Meth.} {\bf#1} (#2) #3.}
\def\np#1#2#3   {{\em Nucl. Phys.} {\bf#1} (#2) #3.}
\def\pcps#1#2#3 {{\em Proc. Cam. Phil. Soc.} {\bf#1} (#2) #3.}
\def\pl#1#2#3   {{\em Phys. Lett.} {\bf#1} (#2) #3.}
\def\prep#1#2#3 {{\em Phys. Rep.} {\bf#1} (#2) #3.}
\def\prev#1#2#3 {{\em Phys. Rev.} {\bf#1} (#2) #3.}
\def\prl#1#2#3  {{\em Phys. Rev. Lett.} {\bf#1} (#2) #3.}
\def\prs#1#2#3  {{\em Proc. Roy. Soc.} {\bf#1} (#2) #3.}
\def\ptp#1#2#3  {{\em Prog. Th. Phys.} {\bf#1} (#2) #3.}
\def\ps#1#2#3   {{\em Physica Scripta} {\bf#1} (#2) #3.}
\def\rmp#1#2#3  {{\em Rev. Mod. Phys.} {\bf#1} (#2) #3.}
\def\rpp#1#2#3  {{\em Rep. Prog. Phys.} {\bf#1} (#2) #3.}
\def\sjnp#1#2#3 {{\em Sov. J. Nucl. Phys.} {\bf#1} (#2) #3.}
\def\spj#1#2#3  {{\em Sov. Phys. JEPT} {\bf#1} (#2) #3.}
\def\spu#1#2#3  {{\em Sov. Phys.-Usp.} {\bf#1} (#2) #3.}
\def\zp#1#2#3   {{\em Zeit. Phys.} {\bf#1} (#2) #3.}

\begin{document}

\sloppy

\pagestyle{empty}

\begin{flushright}
LU TP 95--21 \\
hep-ph/9509264 \\
August 1995
\end{flushright}

\vspace{\fill}

\begin{center}
{\LARGE\bf Low- and High-Mass Components of}\\[2mm]
{\LARGE\bf the Photon Distribution Functions%
$^{\mbox{\normalsize *}}$}\\[10mm]
{\Large Torbj\"orn Sj\"ostrand} \\[2mm]
Department of Theoretical Physics, University of Lund, \\[1mm]
S\"olvegatan 14A, S-223 62 Lund, Sweden \\[2mm]
and \\[2mm]
{\Large Gerhard A. Schuler} \\[2mm]
Theory Division, CERN \\[1mm]
CH-1211 Geneva 23, Switzerland
\end{center}

\vspace{\fill}

\begin{center}
{\bf Abstract}\\[2ex]
\vspace{-0.5\baselineskip}
\noindent
\begin{minipage}{\abstwidth}
Four new parton distributions of the
photon are presented, with a description of theory choices,
properties and applications.
\end{minipage}
\end{center}

\vspace{\fill}

\noindent
\rule[2mm]{50mm}{0.3mm}\\
$^{\mbox{\normalsize *}}$To appear in the Proceedings of the
International Europhysics Conference on High Energy Physics,
Brussels, Belgium, July 27 -- August 2, 1995.\\

\clearpage
\pagestyle{plain}
\setcounter{page}{1}

\section{Introduction}

This is a brief presentation of ref.~1. Related references may be
found on the non-perturbative constraints on the photon structure
function\cite{Gerhard}, on photoproduction\cite{gap} and
on $\gamma\gamma$ physics\cite{Paris}.

There is already available many parton distribution function sets
of the photon\cite{DG,LAC,GRV,GS,AFG,WHIT,FKP}, so why produce more
ones? A few reasons:
\begin{Itemize}
\item The choice of theoretical ansatz is ambiguous, so there is
more room to play than e.g. for the proton.
\item The data are very incomplete and uncertain, so there is a need
for contrasting parametrizations.
\item Compared with previous studies, we put more emphasis on the
subdivision of the distributions into several components of different
physical nature. This is required for a detailed modelling of
$\gamma\p$ and $\gamma\gamma$ hadronic final states, and for a
sophisticated eikonalization approach to the relation between jet
and total cross sections.
\item Our ansatz allows a closed-form extension to the parton
distributions of a (moderately) virtual photon.
\end{Itemize}

\section{Physics Assumptions and Fits}

Photons obey a set of inhomogeneous evolution equations, where the
inhomogeneous term is induced by $\gamma \to \q\qbar$ branchings.
The solution can be written as the sum of two terms,
\begin{equation}
f_a^{\gamma}(x,Q^2) = f_a^{\gamma,\mrm{NP}}(x,Q^2;Q_0^2)
+ f_a^{\gamma,\mrm{PT}}(x,Q^2;Q_0^2) ~,
\end{equation}
where the former term is a solution to the homogeneous evolution
with a (non-perturbative) input at $Q=Q_0$ and the latter is a
solution to the full inhomogeneous equation with boundary condition
$f_a^{\gamma,\mrm{PT}}(x,Q_0^2;Q_0^2) \equiv 0$. One possible
physics interpretation is to let $f_a^{\gamma,\mrm{NP}}$ correspond
to $\gamma \leftrightarrow V$ fluctuations, where
$V = \rho^0, \omega, \phi, \ldots$ is a set of vector mesons,
and let $f_a^{\gamma,\mrm{PT}}$ correspond to perturbative (``anomalous'')
$\gamma \leftrightarrow \q\qbar$ fluctuations. The discrete spectum
of vector mesons can be combined with the continuous (in virtuality
$k^2$) spectrum of $\q\qbar$ fluctuations, to give
\begin{displaymath}
f_a^{\gamma}(x,Q^2) =
\sum_V \frac{4\pi\alphaem}{f_V^2} f_a^{\gamma,V}(x,Q^2) \hspace{30mm}
\end{displaymath}
\vspace{-3mm}
\begin{equation}
\mbox{} + \frac{\alphaem}{2\pi} \, \sum_{\q} 2 e_{\q}^2 \,
\int_{Q_0^2}^{Q^2} \frac{{\d} k^2}{k^2} \,
f_a^{\gamma,\q\qbar}(x,Q^2;k^2) ~,
\end{equation}
where each component $f^{\gamma,V}$ and $f^{\gamma,\q\qbar}$ obeys a
unit momentum sum rule.

Beyond this fairly general ansatz, a number of choices has to be
made, as described in the following.

What is $Q_0$? A low scale, $Q_0 \approx 0.6$~GeV, is favoured if
the $V$ states above are to be associated with the lowest-lying
resonances only. Then one expects $Q_0 \sim m_{\rho}/2$--$m_{\rho}$.
Furthermore, with this $Q_0$ one obtains a reasonable description of
the total $\gamma\p$ cross
section, and continuity e.g. in the primordial $k_{\perp}$ spectrum.
Against this choice speaks worries that perturbation theory may not
be valid at such low $Q$, or at least that higher-twist terms appear
in addition to the standard ones. Alternatively one could
therefore pick a larger value, $Q_0 \approx 2$~GeV, where these
worries are absent. One then needs to include also higher resonances
in the vector-meson sector, which adds some arbitrariness, and one
can no longer compare with low-$Q$ data. We have chosen to prepare
sets for both these (extreme) alternatives.

How handle the direct contribution? Unlike the p, the $\gamma$ has a
direct component where the photon acts as an unresolved probe.
In the definition of $F_2^{\gamma}$ this adds a component
$C^{\gamma}$, symbolically
\begin{equation}
F_2^{\gamma}(x,Q^2) = \sum_{\q} e_{\q}^2 \left[ f_{\q}^{\gamma} +
f_{\qbar}^{\gamma} \right] \otimes C_{\q} +
f_{\g}^{\gamma} \otimes C_{\g} + C^{\gamma} ~.
\end{equation}
Since $C^{\gamma} \equiv 0$ in leading order, and since we stay with
leading-order fits, it is permissible to neglect this complication.
Numerically, however, it makes a non-negligible difference. We
therefore make two kinds of fits, one DIS type with $C^{\gamma} = 0$
and one $\overline{\mrm{MS}}$ type including the universal part
of $C^{\gamma}$\cite{AFG}.

How deal with heavy flavours, i.e. mainly charm? When jet production
is studied for real incoming photons, the standard evolution approach
is reasonable, but with a lower cut-off $Q_0 \approx m_{\c}$ for
$\gamma \to \c\cbar$. Moving to deep inelastic scattering,
$\e\gamma \to \e X$, there is an extra kinematical constraint:
$W^2 = Q^2 (1-x)/x > 4 m_{\c}^2$. It is here better to use the
``Bethe-Heitler'' cross section for $\gamma^* \gamma^* \to \c\cbar$.
Therefore two kinds of output is provided. The parton distributions
are calculated as the sum of a vector-meson part and an anomalous part
including all five flavours, while $F_2^{\gamma}$ is calculated
separately from the sum of the same vector-meson part, an anomalous
part and possibly a $C^{\gamma}$ part now only covering the three
light flavours, and a Bethe-Heitler part for c and b.

Should $\rho^0$ and $\omega$ be added coherently or incoherently?
In a coherent mixture, $\u\ubar : \d\dbar = 4 : 1$ at $Q_0$, while
the incoherent mixture gives $1 : 1$. We argue for
coherence at the short distances probed by parton distributions.
This contrasts with long-distance processes, such as
$\e\gamma \to \e V$.

What is $\Lambda_{\mrm{QCD}}$? The data are not good enough to allow
a precise determination. Therefore we use a fixed value
$\Lambda^{(4)} = 200$~MeV, in agreement with conventional
results for proton distributions.

\begin{figure}[t]
\begin{center}
\mbox{\epsfig{figure=lutp95211.eps,width=150mm}}
\end{center}
\caption{Subdivision of the full $F_2^{\gamma}$ parametrization by
component, compared with data at $Q^2 \approx 5.2$~GeV$^2$.
The total $F_2^{\gamma}$ (except for an overall factor
$\alpha_{\mbox{\tiny em}}$) is shown by the full curve.
The lowest dashed curve gives the VMD contribution, and the next
lowest the sum of VMD and anomalous ones. The third dashed curve,
which coincides with the full curve for the DIS fits, gives
the sum of VMD, anomalous and Bethe--Heitler terms. For the
$\overline{\mrm{MS}}$ fits, the full curve additionally contains the
contribution of the $C^{\gamma}$ term. Note that this last term is
negative at large $x$.}
\end{figure}

In total, four distributions are presented\cite{main}, based on fits
to available data:
\begin{Itemize}
\item SaS 1D, with $Q_0 = 0.6$~GeV and in the DIS scheme.
\item SaS 1M, with $Q_0 = 0.6$~GeV and in the $\overline{\mrm{MS}}$
scheme.
\item SaS 2D, with $Q_0 = 2$~GeV and in the DIS scheme.
\item SaS 2M, with $Q_0 = 2$~GeV and in the $\overline{\mrm{MS}}$
scheme.
\end{Itemize}
Fig.~1 compares these distributions at one $Q^2$ value.

The code for the parametrizations is freely available,
e.g. on the web:\\
{\tt http://thep.lu.se/tf2/staff/torbjorn/lsasgam}

The parametrizations above can be used also for photons with
a virtuality $P^2 \neq 0$. This is done by introducing dipole factors
$m_V^4/(m_V^2 + P^2)^2$ for the VMD components, by changing the lower
cut-off of the anomalous integral from $Q_0^2$ to $\max(Q_0^2, P^2)$,
and by using the proper off-shell expressions for Bethe-Heitler
cross sections etc.

\section{Summary}

Above are presented four new sets of parton distributions and
structure functions of the photon\cite{main}. Owing to the subdivision
of the distributions into different physics components, it is now
possible to construct detailed models of $\gamma\p$ and
$\gamma\gamma$ events. These models are presented
elsewhere\cite{gap,Paris}, with further work in progress.


\begin{thebibliography}{99}

\bibitem{main}
G.A.~Schuler and T.~Sj\"ostrand, CERN--TH/95--62 and
LU TP 95--6, to appear in {\em Zeit. Phys.} {\bf C}.

\bibitem{Gerhard}
G.A.~Schuler, Regensburg preprint TPR--95--15.

\bibitem{gap}
G.A.~Schuler and T.~Sj\"ostrand, \pl{B300}{1993}{169},
{}~\np{B407}{1993}{539}

\bibitem{Paris}
G.A.~Schuler and T.~Sj\"ostrand, in {\em Two-Photon Physics
from DA$\Phi$NE to LEP200 and Beyond}, World Scientific,
Singapore, 1994, Eds. F.~Kapusta and J.~Parisi, p. 163.

\bibitem{DG}
M.~Drees and K.~Grassie,  \zp{C28}{1985}{451}

\bibitem{LAC}
H.~Abramowicz, K.~Charchula and A.~Levy, \pl{B269}{1991}{458}.

\bibitem{GRV}
M.~Gl\"{u}ck, E.~Reya and A.~Vogt, \prev{D46}{1992}{1973}

\bibitem{GS}
L.E.~Gordon and J.K.~Storrow, \zp{C56}{1992}{307}

\bibitem{AFG}
P.~Aurenche, M.~Fontannaz and J.-Ph.~Guillet,
\zp{C64}{1994}{621}

\bibitem{WHIT}
K.~Hagiwara, M.~Tanaka, I.~Watanabe and T.~Izubuchi,
\prev{D51}{1995}{3197}

\bibitem{FKP}
J.H.~Field, F.~Kapusta and L.~Poggioli, \pl{B181}{1986}{362},
{}~\zp{C36}{1987}{121}

\end{thebibliography}
\end{document}